\documentstyle[epsfig]{elsart}
\def\grsim{\mathrel{\hbox{\lower1ex\hbox{\rlap{$\sim$}\raise1ex\hbox{$>$}}}}}
\def\losim{\mathrel{\hbox{\lower1ex\hbox{\rlap{$\sim$}\raise1ex\hbox{$<$}}}}}
\makeatletter
\def\ps@copyright{\let\@mkboth\@gobbletwo
  \def\@oddhead{}%
  \let\@evenhead\@oddhead
  \def\@oddfoot{\small\sl
      GSI-Preprint 95-68, accepted for publication in Nucl. Phys. A
      \hfill }
  \let\@evenfoot\@oddfoot}
\begin{document}
\begin{frontmatter}
\title{
Photoproduction of vector mesons off nucleons near threshold}
\author{Bengt Friman}
\address{
        GSI, Postfach 11 05 52\\
        D-64220 Darmstadt, Germany\\
        Institut f\"ur Kernphysik, TH Darmstadt\\
        D-64289 Darmstadt, Germany}
\and
\author{Madeleine Soyeur}
\address{Commissariat \`a l'Energie Atomique\\
        Laboratoire National Saturne\\
        CE de Saclay\\
        F-91191 Gif-sur-Yvette Cedex, France}

\begin{abstract}
We propose a simple meson-exchange model of the photoproduction of
$\rho$- and $\omega$-mesons off protons near threshold $(E_\gamma \losim 2
\, GeV)$. This model provides a good description of the
available data and implies a large $\rho$-nucleon interaction in the
scalar channel ($\sigma$-exchange). We use this phenomenological
interaction to estimate the leading contribution to the self-energy
of $\rho$-mesons in matter. The implications of our
calculation
for experimental studies of the $\rho$-meson mass in nuclei
are discussed.
\end{abstract}
\end{frontmatter}
\clearpage

\section{Introduction}

The properties of vector mesons in the nuclear medium appear to be
important not only for the understanding of nuclear dynamics but also
possibly to reveal a nonperturbative aspect of quantum chromodynamics
(QCD), the restoration of chiral symmetry at high baryon density or
temperature [1-3].

As a consequence of the spontaneous breaking of chiral symmetry, the
QCD vacuum is characterized by a non-zero expectation value, the quark
condensate $< \bar{q}q >$. When the baryon density or temperature
increases, this expectation value decreases and would vanish at
the critical density or temperature, if chiral symmetry was not
explicitly broken by finite quark masses. We consider here the
decrease of the quark condensate with increasing density. In the low
density approximation, expected to be valid for densities below the
saturation density ($\rho_0$), it has been shown that this decrease
is linear as a consequence of the Feynman-Hellmann theorem
\cite{4,5}. The slope is such that the value of  $< \bar{q}q >$ at $\rho_0$
is about 30\% smaller than at $\rho=0$. Observing this effect
experimentally is a major challenge.

Brown and Rho \cite{1,6} suggested that the masses of vector mesons
($\rho$- and $\omega$-mesons) are generated by their repulsive
coupling to the quark condensate and therefore decrease with
increasing baryon density. In their approach, vector meson masses scale
with the cubic root of the quark condensate and have dropped by about
100 MeV at $\rho_0$. Decreasing in-medium vector meson masses linked
to the decrease of the quark condensate were also obtained by making
use of the techniques of QCD sum rules in matter \cite{2,3}. However,
this result
relies on the factorization of the 4-quark condensate, a possibly
questionable approximation. Experimental studies of the in-medium
vector meson masses are therefore very much needed and could shed some
light on the origin of hadron masses and their relation to chiral
symmetry.

The observation of the in-medium behaviour of vector mesons requires
processes in which vector mesons are produced and decay inside
nuclei. The candidates for such measurements are necessarily broad
mesons, as their mean free path has to be shorter that the nuclear
size. This is the case for the $\rho$-meson which has a total width
$\Gamma_\rho$ of $(151.2 \pm 1.2) $ MeV
and a propagation length $c\tau _\rho$ of $ \sim 1.3$ fm
\cite{7}. The small free space width \cite{7} of the $\omega$-meson
[$\Gamma_\omega = (8.43 \pm 0.10)$ MeV and $c \tau_\omega \simeq 23.4$
fm] precludes such studies, unless $\omega$-mesons acquire a large
width in matter. It may nevertheless be possible to study the in-medium
properties of $\omega$-mesons in the particular recoilless kinematics where
they are produced approximately at rest in nuclei \cite{8}. We focus
on $\rho$-mesons and consider their production in nuclear
targets close to threshold, where their formation time and lifetime
are not dilatated by large Lorentz factors. The spectrum of the
corresponding vector excitation can be best studied through its
$e^+e^-$ decay, as the emitted leptons are not distorted by strong
interactions. Such experiments are planned at CEBAF \cite{9} and GSI
\cite{10}.

The main purpose of this work is to construct a model for the
photoproduction
of $\rho$- and $\omega$-mesons off protons near threshold
($E_\gamma \losim 2 \, GeV$) as a first step to study later their
photoproduction in nuclei in that kinematic regime. We
show that, close to threshold, the $\gamma p \rightarrow \omega p$
and the $\gamma p \rightarrow \rho^0 p$ cross sections \cite{11}
can be well described by a simple meson-exchange model. In this
picture, the $\omega$ photoproduction appears dominated by
$\pi$-exchange while the $\rho$ photoproduction cross section is
given mainly by $\sigma$-exchange. Assuming the Vector Dominance
of the electromagnetic current \cite{12},
we can extract from this last process a phenomenological
$\rho$-nucleon interaction. We use it to derive the
leading contribution to the $\rho$-meson self-energy in matter.
The $\rho$-nucleon interaction in the scalar
channel leads to a $\rho$-meson self-energy in matter which is large
and attractive. We discuss the theoretical interpretation of this
effect and its consequences for experimental studies of the $\rho$-meson
mass in nuclei.

We present our model for the $\gamma p \rightarrow \omega p$ and
$\gamma p \rightarrow \rho^{0}p$ reactions
near threshold in Section 2. The $\rho$-meson self-energy in matter
generated to lowest order by the $\rho$-nucleon interaction
deduced from this model with the $\sigma \rho \rho$ coupling
we consider
is calculated and discussed in Section 3. We conclude
by a few remarks in Section 4.

\section{Vector meson photoproduction off protons near threshold}

The production of the $\rho$(770)- and $\omega$(782)-mesons induced
by photons of a few GeV scattered from proton targets has been studied
extensively \cite{13}. At sufficiently high energy, typically $E_\gamma >
3 \, GeV$, this process appears diffractive: the total cross section is
roughly energy independent, the differential cross section
$d\sigma/dq^2$ exhibits a sharp peak in the forward direction, the
t-channel exchange has the quantum numbers of the vacuum
(Pomeron exchange) and the amplitude is mostly imaginary.

At lower energies $(E_ \gamma < 2 \, GeV)$, meson-exchange contributions
are expected to play a role. The importance of the $\pi$-exchange in
the $\gamma p \rightarrow \omega p$ reaction near threshold was
emphasized long ago and is supported by data taken with linearly
polarized photons \cite{14}.

Following the early suggestion of Berman and Drell \cite{15} and work by
Joos and Kramer \cite{16}, we describe the $\gamma p \rightarrow V
p \ (V = \rho^0, \omega)$ reaction near threshold by t-channel exchanges
and assume that the production amplitudes are dominated by the
exchange of light mesons. There is no rigorous
justification for this choice. We favour this model because it
provides a reasonable description of the data with a small number
of parameters. It is in particular much simpler than the s-channel
picture, based on Compton-like graphs, which requires summation over a
very large number of overlapping baryon resonances \cite{17}.

For reasons which will become clear later, we begin our discussion by
the $\gamma p \rightarrow \omega p$ reaction. In the Born
approximation, only scalar- and pseudoscalar-meson exchanges can
contribute to our t-channel description. Vector-meson exchanges are
not allowed by charge conjugation invariance. In the pseudoscalar
channel, the $\eta$-exchange contribution is small
compared to the $\pi$-exchange for two reasons. First, the total
$\gamma p \rightarrow \omega p$ cross section is dominated by the low
$q^2$ regime, where the $\eta$-exchange is strongly suppressed compared to the
$\pi$-exchange because of their large mass difference. Second,
the $\omega \rightarrow \eta \gamma$ branching
ratio is about two orders of magnitude smaller than the $\omega
\rightarrow \pi^o \gamma$ branching ratio \cite{7}. An explicit calculation
shows that the
ratio of the coupling constants $g^2_{\omega \pi \gamma} / g^2_{\omega
\eta \gamma}$ is of the order of 15. This implies a further
suppression of the
$\eta$-exchange compared to the $\pi$-exchange. Consequently, we are
left effectively with two contributions, the $\pi^0$-exchange and the
$\sigma$-exchange shown in Fig.~1.

The relative importance of these two contributions may be assessed by
studying
the radiative transitions of $\omega$-mesons to neutral pions and
$\sigma$'s. Since the $\sigma$-meson is an effective degree of freedom
associated with the propagation of two pions in a relative s-wave
state, a first
estimate of the strength of these transitions can be obtained from the
branching ratio of the $\omega$-meson into the $\pi^0 \gamma$ and
$\pi^+\pi^- \gamma$ channels (in the latter case, the coupling of the
two pions in a $\rho^0$-like state is forbidden by charge conjugation
invariance). The $\omega \rightarrow \pi^0 \gamma$ branching ratio
is known to be of the order of 8.5 \% \cite{7}. Unfortunately, the only
data available on the $\omega \rightarrow \pi^+ \pi^- \gamma$
branching ratio is an upper limit of 3.6 x 10$^{-3}$ \cite{7}. This
suggests nevertheless that the $\pi$-exchange is the dominant
contribution. Another approach to evaluate these vertices is Vector Meson
Dominance \cite{12}. Because of isospin conservation, the $\pi^0\omega$ channel
will couple only to the isovector part of the electromagnetic current
while the $\sigma \omega$ channel will couple solely to its isoscalar
part. Using the current-field identities of Kroll, Lee and Zumino
\cite{18}, we identify the isoscalar and isovector parts of the
electromagnetic current with the $\omega$- and $\rho$-meson currents
respectively. This is shown pictorially in Fig. 2. The relation between
the electromagnetic current and the vector fields is given by \cite{18}
\begin{equation}
{\cal J}_\mu^{em} (I=1) = \frac{e M_\rho^2} {2g_\rho} \rho_\mu ,
\end{equation}
\begin{equation}
{\cal J}_\mu^{em} (I=0) = \frac{e M^2_\omega}{2g_\omega} \omega_\mu,
\end{equation}
where $M_\rho$ and $M_\omega$ are the $\rho$ and $\omega$-masses and
$g_\rho$ and $g_\omega$ are dimensionless constants, which can be
determined from the $e^+e^-$ partial decay widths of the $\rho$- and
$\omega$-mesons \cite{7} to be
\begin{equation}
g^2_\rho = 6.33
\end{equation}
and
\begin{equation}
g^2_\omega = 72.71.
\end{equation}
Because the $\pi^0$- and $\sigma$-mesons have opposite parity, the
$\pi$- and $\sigma$-exchange amplitudes do
not interfere. Consequently, the ratio of the $\pi$- to $\sigma$-exchange
contributions to the total $\gamma p \rightarrow \omega p$ cross
section is proportional to $g^2_\omega/g^2_\rho$,
indicating again the dominance
of the $\pi$-exchange term. This is consistent with the
experimental observation that the exchange of unnatural parity
states in the t-channel dominates the $\gamma p \rightarrow \omega p$
cross section near threshold \cite{14}. We calculate first
the $\pi$-exchange contribution to this reaction.

The $\pi$NN vertex is described by the pseudoscalar coupling
\begin{equation}
{\cal L} ^{int} _{\pi NN} = - i g_{\pi NN} \bar{N} \gamma_5 (\vec{\tau}\cdot
 \vec{\pi}) N
\end{equation}
with a dipole form factor,
\begin{equation}
F_{\pi NN} = \frac {\Lambda^2_\pi - m^2_\pi} {\Lambda^2_\pi - q^2},
\end{equation}
where $m_\pi$ is the pion mass, $g_{\pi NN}$ the pion-nucleon coupling
constant (we take $ g^2_{\pi NN}/{4\pi}=14$) and $\Lambda_\pi$ the
cut-off characterizing the pion-nucleon vertex. Lorentz and gauge
invariances imply that the $\omega\pi\gamma$ vertex is of the form,
\begin{equation}
{\cal L}^{int}_{\omega\pi^0 \gamma} =
e \, \frac{g_{\omega\pi\gamma}}{M_\omega} \, \varepsilon_{
\alpha\beta\gamma\delta} \, \partial^\alpha A^\beta \, \partial^\gamma
\omega^\delta \pi^0,
\end{equation}
where $\varepsilon_{\alpha \beta \gamma \delta}$ is the totally antisymmetric
Levi-Civita tensor and the dimensional coupling strength has been normalized
to the $\omega$-meson mass. We use the experimental partial decay width
for $\omega \rightarrow \pi^0 \gamma$ \cite{7} to determine the coupling
constant $g^2_{\omega\pi\gamma}$ = 3.315.
Using the current-field identity
(1), we can calculate the $\omega\pi^0\gamma$ vertex assuming
$\rho$-dominance (as shown in the left graph of Fig. 2) and express the
$\omega\rho^0\gamma$ vertex as a function of the
$\omega \rho^0 \pi^0$ interaction. Writing by analogy to (7)
\begin{equation}
{\cal L} ^ {int}_ {\omega \pi^0 \rho^0} =
\frac{g_{\omega\pi\rho}}{M_\omega} \,
\varepsilon_{\alpha\beta\gamma\delta} \, \partial^\alpha \rho^{0\beta}
\partial^\delta \omega^\rho \pi^0,
\end{equation}
we obtain
\begin{equation}
g^2_{\omega\pi\rho} = 4 \, g^2_\rho \, g^2_{\omega\pi\gamma} \simeq 84.
\end{equation}
We fix the non-locality of this vertex from the study of the $\omega
\rightarrow \pi^0\mu^+\mu^-$ form factor analyzed with the interaction
Lagrangian (8) and the current-field identity (1).
This model of the vertex is shown in Fig. 3 and in Fig. 4 we display the
result of the calculated form factor, with a local $\pi\rho\,\omega$
vertex (dashed curve) and with an extended vertex of the form
$\Lambda^2_\rho/(\Lambda^2_\rho-q^2)$ with $\Lambda_\rho = M_\rho$ (full
curve).
It is clear
that a simple Vector Dominance Model with a local $\pi\rho\,\omega$
vertex does not describe the data \cite{19,20} and that an additional form
factor is required. We shall not discuss here the theoretical
understanding of this vertex \cite{21,21a}, which is closely related to
the non-Abelian anomaly \cite{22}. We assume that the mass scale of
the $\pi\rho\,\omega$ form factor obtained by studying the Dalitz
decay of the $\omega$-meson characterizes the size of the
$\pi\rho\,\omega$ vertex. Consequently, we employ the same value
of the cut-off
to describe the dependence of the form factor on the pion four-momentum
when the $\rho$-meson is on-shell and the pion is off-shell.

In this way, all the parameters of our $\pi$-exchange model of the
$\gamma p \rightarrow \omega p$ cross section are fixed, except for
$\Lambda_\pi$, the cut-off mass in the $\pi$NN form factor. We
determine $\Lambda_ \pi$ from a fit to the total
$\gamma p \rightarrow \omega p$ cross section near threshold
$(E^{thr}_\gamma$ = 1.108 GeV), knowing from data that the domain of
validity of this pion-exchange model should not extend beyond $\sim$ 2
GeV, as natural parity exchanges (Pomerons) are expected to be
responsible for a large fraction of the cross section at $E_\gamma$ = 2.8 GeV
\cite{14}. We find that the best fit is obtained for $\Lambda_\pi$ = 0.7
GeV. The total cross section for the $\gamma p \rightarrow \omega p$
reaction obtained in our model is compared to
data \cite{11} in Fig. 5. The $\pi$-exchange model accounts
for the data below 2
GeV, while for larger photon energies, an additional contribution is
needed. The missing cross section is almost independent of energy,
a behaviour consistent with
the Pomeron-exchange term
becoming dominant. The value of $\Lambda_\pi$ = 0.7 GeV
is consistent with analyses of pion-nucleon data \cite{23}
and with recent work indicating that the larger
$\Lambda_\pi$ obtained in models of the nucleon-nucleon interaction
can be reduced if the correlated $\rho\pi$-exchange is explicitly
included \cite{24}.

To have a more stringent test of the form factors, we have calculated
with the $\pi$-exchange model discussed above the differential cross
section $d\sigma/dq^2$ for the $\gamma p \rightarrow \omega p$ reaction
induced by photons of energies 1.4 $<\mbox{E}_\gamma<$1.8 GeV. The
expression for $d\sigma/dq^2$ is simply
\begin{eqnarray}
\frac{d\sigma}{dq^2}^{\gamma p \rightarrow \omega p} = \alpha\,
\frac{g^2_{\pi\rho\,\omega}}{4\pi} \, \frac{g^2_{\pi pp}}{4 \pi}
\, \frac{\pi^2}{4g^2_\rho} \ \frac{(\hbar c)^2}{M^2_\omega}\,
\frac{1}{E^2_\gamma} \,
\frac{-q^2}{4 M ^2_p}  \left[ \frac{M^2_\omega - q^2}{m^2_\pi
- q^2} \right]^2 \nonumber \\
 \left[ \frac{\Lambda^2_\pi - m^2_\pi}{\Lambda^2_\pi - q^2} \right]^2
 \left[  \frac{M^2_\rho-m^2_\pi}{M^2_\rho - q^2} \right]^2
\end{eqnarray}
and the result of the calculation compared to the data \cite{11} is
displayed in Fig. 6.

Within error bars, the general trend of the $q^2$-dependence is well
reproduced for $|q^2|  < 0.5\, GeV^2$. Beyond this value, which
corresponds to the range of our cut-offs, we do not expect our
model to be valid.

We conclude from this discussion that, sufficiently close to threshold
$(E_\gamma < 2 $ GeV), there is no clear indication of anything else
contributing to the $\gamma p \rightarrow \omega p$ process in the
t-channel but the $\pi$-exchange. It is also remarkable that for values
of the coupling constants and the cut-offs consistent with other
processes, the total cross section is correctly given by this simple
model.

Motivated by this result, we study the
$\gamma p \rightarrow \rho^0 p$ reaction within the same approach.
As in the case of $\omega$
photoproduction, vector-meson exchanges
are not allowed and we consider only the $\pi^0$- and
$\sigma$-exchanges shown in
Fig. 7. Only the isoscalar part of the
electromagnetic current ($\omega$ field) contributes
to the $\pi$-exchange graph, while
the $\sigma$-exchange selects the isovector part
($\rho^0$ field).

In contrast to what happens in the
$\omega$ photoproduction, the scalar-exchange
appears to be the dominant contribution to the $\gamma p \rightarrow
\rho^0 p$ reaction. We see two reasons for this. Information on the
scalar and pseudoscalar coupling strengths can be inferred from
an analysis of the $\rho^0$ radiative decays. In spite of
unfavourable phase space,
the $\rho^0 \rightarrow \pi^+ \pi^- \gamma$ branching
ratio of (9.9 $\pm$ 1.6) x 10$^{-3}$ is an order of magnitude
larger than the $\rho^0 \rightarrow \pi^0 \gamma$ branching
ratio of
(7.9 $\pm$ 2.0) x 10$^{-4}$ \cite{7} (again the $\rho^0$-like
coupling of the $\pi^+ \pi^-$ pair in the $\rho^0 \rightarrow \pi^+
\pi^- \gamma$ decay is forbidden by charge conjugation invariance). Using
the current-field identities (1)-(2), it is also clear from
Fig. 7 that the $g^2_\omega/g^2_\rho$ ratio of $\sim$10 is now in favour of
the $\sigma$-exchange contribution to the cross section.

We keep both $\pi$- and $\sigma$-exchanges in the calculation of the
$\gamma p \rightarrow \rho^0 p$ cross section. The $\pi$-exchange
contribution is evaluated with the same parameters as those used for
the description of the $\gamma p \rightarrow \omega p$ cross
section. The $\sigma$-exchange involves two vertices. We describe the
$\sigma$NN vertex by the scalar coupling,
\begin{equation}
{\cal L} ^ {int}_{\sigma NN} = g _ {\sigma NN} \bar N N \sigma,
\end{equation}
with the form factor
\begin{equation}
F_{\sigma NN} = \frac {\Lambda^2_\sigma - m^2_\sigma}
{\Lambda^2_\sigma - q^2} .
\end{equation}

The $\sigma$ mass $m_\sigma$ is taken to be 500 MeV as the
$\sigma$-meson introduced here should be viewed as representing the
exchange of two uncorrelated as well as two resonating
pions. For the coupling constant, we use the standard value
$g^2_{\sigma NN} / 4 \pi = 8$ \cite{25}. The cut-off $\Lambda _\sigma$ will be
determined later. By analogy to the $\omega \pi^0 \rho^0$ vertex, we
construct first the $\rho^0 \sigma \rho^0$ vertex guided by the
current-field identities (1)-(2),
\begin{equation}
{\cal L} ^{int}_{\rho^0 \sigma \rho^0} = \frac {g_{\sigma \rho
\rho}}{M_\rho}   \left[ \partial^\alpha \rho^{0\beta} \partial_\alpha
\rho^0_\beta - \partial^\alpha \rho^{0\beta} \partial_\beta
\rho^0_\alpha \right] \sigma,
\end{equation}

and use a $\sigma\rho\rho$ form factor of the monopole form
${(\Lambda^2_{\sigma\rho\rho}-m^2_\sigma)}/
{(\Lambda^2_{\sigma\sigma\rho} - q^2)}$.
We then fit the values of $\Lambda_\sigma$, $\Lambda_{\sigma\rho\rho}$
and
$g_{\sigma \rho \rho}$ to data.

The important constraints on the vertices come from the differential
cross section $d \sigma/ dq^2$ given as the sum of
$\pi$- and $\sigma$-exchanges by
\newpage
\begin{eqnarray}
\frac {d \sigma ^ {\gamma p \rightarrow \rho^0 p}}{dq^2} = \alpha \,\frac{g^2_
{\pi \rho \omega}}{4 \pi} \, \frac {g^2_ {\pi pp}} {4 \pi} \,
\frac {\pi^2} {4 g^2_\omega} \, \frac {(\hbar c)^2}{M^2_\omega} \,
\frac {1}{E^2_\gamma}\,
\frac{-q^2}{4  M^2_p}  \left[ \frac  {M^2_\rho - q^2} {m^2_\pi - q^2}
\right]^2  \left[ \frac  {\Lambda^2_\pi - m^2_\pi}{\Lambda ^2_ \pi - q^2}
\right]^2 \nonumber \\ \left[ \frac {M^2_\rho-m^2_\pi}
{M^2_\rho - q^2} \right]^2
+ \alpha\, \frac {g^2_{ \sigma \rho \rho}}{ 4 \pi} \,
\frac {g^2_{ \sigma p p}}
{4 \pi} \ \frac {\pi^2} {4 g^2_\rho} \, \frac {(\hbar c)^2} {M^2_\rho}
\, \frac {1} {E^2_\gamma}\,
\frac {4 M ^2 _ p - q^2} {4 M^2 _ p} \left[ \frac {M^2_\rho -
q^2}
{m^2_ \sigma - q^2} \right] ^2
\nonumber \\ \left[ \frac {\Lambda^2_\sigma -
m^2_\sigma}{\Lambda ^2 _\sigma - q^2} \right]^2
\left[ \frac
{\Lambda^2_{\sigma\rho\rho}- m^2_\sigma} {\Lambda^2_{\sigma\rho\rho}
- q^2} \right]^2.
\end{eqnarray}

We find that the total $\gamma p \rightarrow \rho p$ cross section for
$E_\gamma <$ 2 GeV and the differential cross section $\d
\sigma/dq^2$ for 1.4 $< E_\gamma <$ 1.8 GeV are well reproduced for
$\Lambda _ \sigma$ = 1 GeV, $\Lambda _ {\sigma\rho\rho}$
= 0.9 GeV and ${g^2_{\sigma \rho \rho}}/{4 \pi}$ =
14.8. The calculated curves are compared to data \cite{11} in Figs. 8 and
9.

The data points for the total cross section are very much spread,
reflecting the uncertainties in extracting the $\rho$-meson production
cross section from the $\pi^+ \pi^-$ pair emission \cite{11}. Better data
are very much needed. In view of this, there is a  large uncertainty
in the fit. Nevertheless, it is clear
that our $(\pi + \sigma)$-exchange model is below
the data for $E_\gamma > $ 2 GeV, as
expected. It would be important to determine the $\rho$
photoproduction cross section between 1.5 and 2 GeV better to find out
where it becomes
diffractive. We should remark that we use
a zero-width description of the produced $\rho$-meson. This is in
general a good approximation for the quantities we calculate, in
particular since
we do not consider cross sections in a
specific $\rho$ decay channel. However, in the zero-width
approximation, we cannot reproduce
the lowest energy point $(E _\gamma$ = 1.05 GeV), which is
below ``threshold''.

As shown in Fig. 9, the $\pi$-exchange contributes very modestly to the
cross section, at any momentum transfer.
The differential cross
section provides definite constraints on the cut-offs and couplings for
the $\sigma$-exchange.
In the total cross section, there is a trade-off between the form factors
and the $\sigma\rho\rho$ coupling strength. However, increasing the cut-off
masses to compensate for a smaller value of $g_{\sigma\rho\rho}$ would
produce a differential cross section which is too flat.

The $\rho^0 \sigma \rho^0$ vertex (13), suggested by current-field
identities, is not unique. Lorentz invariance
allows also for a nonderivative coupling of the form,

\begin{equation}
{\cal L} ^{int}_{\rho^0 \sigma \rho^0} = \frac {1} {2} \tilde{g} _
{\rho \rho \sigma} M_\rho \rho ^\mu \rho_\mu  \sigma.
\end{equation}

This vertex used in the calculation of the photoproduction
cross section leads however to a result which is not gauge invariant.	
Hence, we do not consider it in this work.

The phenomenological
T-matrix for the $\rho$-nucleon interaction, obtained in the Born
approximation, fits the available
data on the $\gamma p \rightarrow \rho^0 p$ reaction near
threshold. In the following section we employ
this interaction to derive the $\rho$-meson
self-energy in the nuclear medium.

\section{The $\rho$-meson self-energy in matter}

To leading order, the $\rho$-meson self-energy in nuclear matter is
given by the tadpole diagram shown in Fig. 10.
This is, in the Born approximation, nothing but the low-density
theorem discussed in Ref. \cite{26}.
We calculate this term
with $\sigma \rho^0$ coupling
given by Eqs. (13). The corresponding
self-energy is denoted by $\Sigma_\rho$. We find
\begin{equation}
\Sigma_\rho (q^2) = - \frac {g_{\sigma \rho \rho}}{M_\rho} \,
\frac {g_{\sigma N N}} {m^2_\sigma} \,
\frac {\Lambda^2_\sigma - m^2_\sigma}{\Lambda^2_\sigma} \,
\frac {\Lambda^2_\rho - m^2_\sigma}{\Lambda^2_\rho}\,
q^2 \rho ,
\end{equation}
where q is the $\rho$-meson 4-momentum and $\rho$ is the matter
density.
Using the parameters obtained in the previous section, we get
\begin{equation}
\Sigma_\rho (q^2) \simeq - 2.81 \, q^2 \, \rho (fm^{-3})\,  GeV^2.
\end{equation}
We have assumed that the a priori unknown phase of the
$g_{\sigma \rho \rho}$ coupling constant is such that the corresponding
self-energy is attractive.
This follows
naturally if we think of the $\sigma$-exchange as
an effective $2\pi$-exchange process where the intermediate states are
dominated by states of
energies higher then the $\rho$ mass. An example of such a process
with an intermediate $\omega$ particle-hole state is shown in Fig. 11b.

The corresponding in-medium inverse $\rho$-meson propagator is
\begin{equation}
D^{-1}_\rho (q^2) = D^{0\,-1}_\rho - \Sigma_\rho,
\end{equation}
with $D^{0\,-1}_\rho = q^2 - M^2_\rho$. From Eqs. (17), we obtain
\begin{equation}
D_\rho ^{-1} (q^2) \ = q^2  \left[ 1 + 2.81 \,\rho (fm^3)
\right] - M^2_\rho.
\end{equation}

With the derivative $\sigma \rho$ coupling of Eq. (13), the self-energy
gives rise to a wave function renormalization. Equation (19) can indeed
be written as

\begin{equation}
D_\rho(q^2) = {Z\over q^2 - Z M_\rho^2 + i \varepsilon} \label{proprho}
\end{equation}

with

\begin{equation}
Z^{-1} = 1 + \frac{g_{\sigma \rho \rho}}{M_\rho} \frac{g_{\sigma N
N}}{m_\sigma^2} \frac{\Lambda_\sigma^2-m_\sigma^2}{\Lambda_\sigma^2}
\rho = 1 + 2.81 \rho \, (fm^{-3}).
\end{equation}

At $\rho = \rho_0$, the pole is shifted down from $M_\rho$ to 0.82
$M_\rho$ and the strength is reduced by 32 \%. This result may
at first sight seem dubious, since the total strength satisfies a sum rule

\begin{equation}
\int {dq^2\over \pi} \left[ - Im D_\rho(q^2)\right] = 1.\label{sumrule}
\end{equation}

This can be understood by considering the underlying
picture where the $\sigma$-exchange effectively accounts for a $2
\pi$-exchange process (see Fig. 11a). The associated $\rho$-meson
self-energy (Fig. 11b) is energy-dependent and has an imaginary
part. In the case illustrated in Fig. 11b, it corresponds
to the decay of a $\rho$-meson (off-shell) into
an $\omega$-meson and a particle-hole pair. This physical process also
shows up in the analytical structure of the $\rho$ propagator as a cut in the
complex $q^2$ plane. By replacing the $2 \pi$-exchange effectively by
a $\sigma$-exchange, one implicitly assumes that this cut is far away
from the energies of interest, so that the energy dependence due to
the intermediate states can be neglected. For the problem at
hand, this assumption is not rigorously justified. In what follows,
we shall however assume that the dominant contributions
to the intermediate states have masses larger than the
$\rho$-meson mass.

To see how this picture affects the
sum rule, let us consider the following schematic model. Replace the
cut in the self-energy by a simple pole located at $q^2 = M^2$. The
$\rho$-meson self-energy is then of the form

\begin{equation}
\Sigma_\rho(q^2) = f^2 q^2 {M^2\over q^2 - M^2}. \label{self}
\end{equation}

We assume derivative coupling so that the self-energy is proportional
to $q^2$. Thus, the self-energy (23)
reduces to the $\sigma$-induced self-energy (16) in the limit $M
\rightarrow \infty$.

The propagator

\begin{equation}
D_\rho(q^2) = {1\over q^2 - M_\rho^2 - \Sigma_\rho (q^2)}\label{prop}
\end{equation}

has two poles, a
low mass one $q_A^2$ near $M_\rho^2$, and a high mass one $q_B^2$ near
$M^2$. The low mass pole is identified with the in-medium $\rho$-meson.
For $M >> M_\rho$, one finds, to leading order in $(M_\rho/M)^2$,
$q_A^2 = M_\rho^2/(1+f^2)$ and  $q_B^2 = (1+f^2) M$. The corresponding
residues are $Z_A = 1/(1+f^2)$ and $Z_B = f^2/(1+f^2)$ and the sum
rule is fulfilled, i.e., $Z_A + Z_B = 1$. In the limit $M \rightarrow
\infty$, $q_B^2 \rightarrow \infty$ but $Z_B$ remains finite.
The missing strength is moved out to $q^2 = \infty$.
Consequently, the propagator (20) does not satisfy the sum
rule because it is valid only at low energies,
while the sum rule involves all energies. Thus, our choice of phases for the
coupling constants, which leads to a reduced $\rho$-mass in medium, is
consistent also with the sum rule (22).

\section{Concluding remarks}

We have shown in this paper that the presently available
data on the photoproduction of $\rho$- and $\omega$-mesons
off protons near threshold (E$_\gamma <$ 2 GeV)
can be well described by a simple
one-boson-exchange model. The dominant processes are the
pseudoscalar $\pi$-exchange for the $\omega$ photoproduction
and the scalar exchange, which we parametrize by an effective
$\sigma$-meson, for the $\rho^0$ photoproduction.

The $\pi$-exchange model of the $\gamma p \rightarrow \omega p$
is constructed using parameters constrained by other data with
moderate extrapolations. We rely however on the measurement
of the $\omega \rightarrow \pi^0 \mu^+ \mu^-$ form factor
for values of the invariant $\mu^+ \mu^-$ pair mass of the
order of 500 to 600 MeV (close to the $\rho$ pole). These data
\cite{19,20} have large error bars and have been obtained with
very few events. A better measurement of this form factor
appears very important for the understanding of the
$\omega \pi^0 \gamma^*$ vertex, in which $\gamma^*$ is a
virtual time-like photon. It is interesting that both
our $\pi$-exchange model and polarization data \cite{14}
on the $\gamma p \rightarrow \omega p$ reaction
close to threshold do not seem to require a scalar exchange.
This suggests that the $\sigma \omega$ interaction is
substantially weaker
than the $\sigma \rho$ interaction.

The $(\pi + \sigma)$-exchange model of the
$\gamma p \rightarrow \rho^0 p$
reaction is much less constrained and it
is clear that it
does not provide a detailed understanding of the photoproduction of
$\rho$-mesons near threshold.
It is indeed desirable to build a
dynamical model of the effective $\sigma$-exchange, going
beyond the Born approximation, to unravel the
physics of this process.
This will make it possible to determine the important contributions
to the intermediate states of the type shown in Fig. 11b and put our
discussion of Section 3 on a firmer basis. Such a dynamical model
would also allow a consistent discussion of the $\sigma \rho$
and $\sigma \omega$ vertices. It could provide a more detailed
comparison to data through a study
of the energy-dependence of the ratio of the real to the imaginary
parts of the production amplitude, a quantity expected to be very
sensitive to the underlying dynamics. Work in this direction is in
progress. We also reemphasize the need for much more accurate data than those
presently available \cite{11}.

Our discussion of the $\rho$-meson self-energy in matter
shows the importance of the structure of the $\sigma\rho$ vertex.
The off-shell behaviour plays a very important role in determining
this self-energy from the $\rho^0$ photoproduction reaction.
Additional constraints on the $\sigma\rho$ vertex from
different kinematic regimes (available for example in the electroproduction
of vector mesons near threshold) would be very important
to establish our result on a more solid basis.
Systematic studies of processes involving vector mesons
in a consistent theoretical framework, such as nonlinear
chiral Lagrangians in which vector mesons are introduced as
dynamical gauge bosons \cite{27}, could also provide useful guidance
for choosing the form of the interaction vertices.
The particular form of the $\rho$-meson self-energy will affect
the dilepton spectrum of $\rho$-like excitations in nuclei. The
theoretical interpretation of this spectrum is therefore closely
linked to the understanding of the momentum dependence of the $\rho$
self-energy at finite baryon density.

\section{Acknowledgements}

We thank George Bertsch, Mannque Rho and Wolfram Weise
for very helpful discussions. One of us (M. S.) acknowledges the generous
hospitality of GSI, where part of this work was done.

\vfill\eject

\newpage
\setlength{\unitlength}{1mm}
\begin{picture}(150,80)
\put(0,10){\epsfig{file=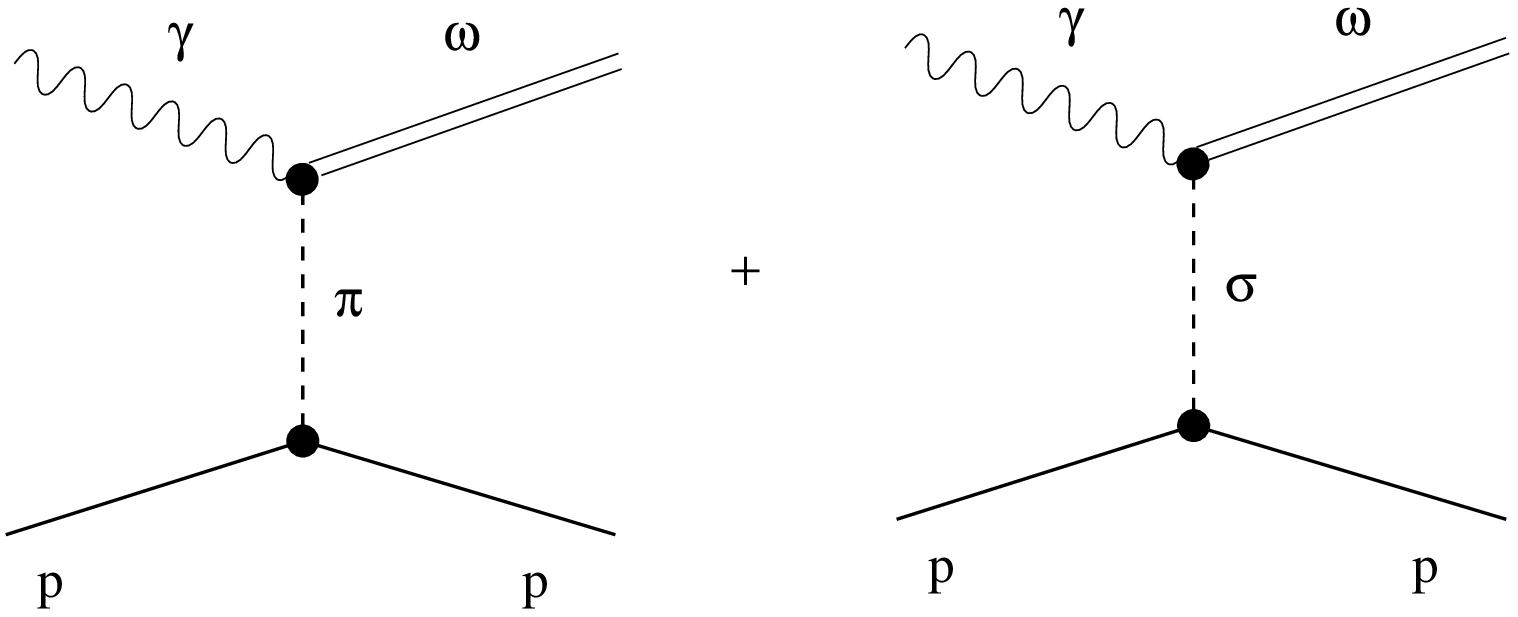,
height=60mm,angle=-90}}
\end{picture}

Fig.~1 t-channel exchange contributions to the $\gamma p \rightarrow
\omega p$ reaction.

\setlength{\unitlength}{1mm}
\begin{picture}(150,80)
\put(-10,-10){\epsfig{file=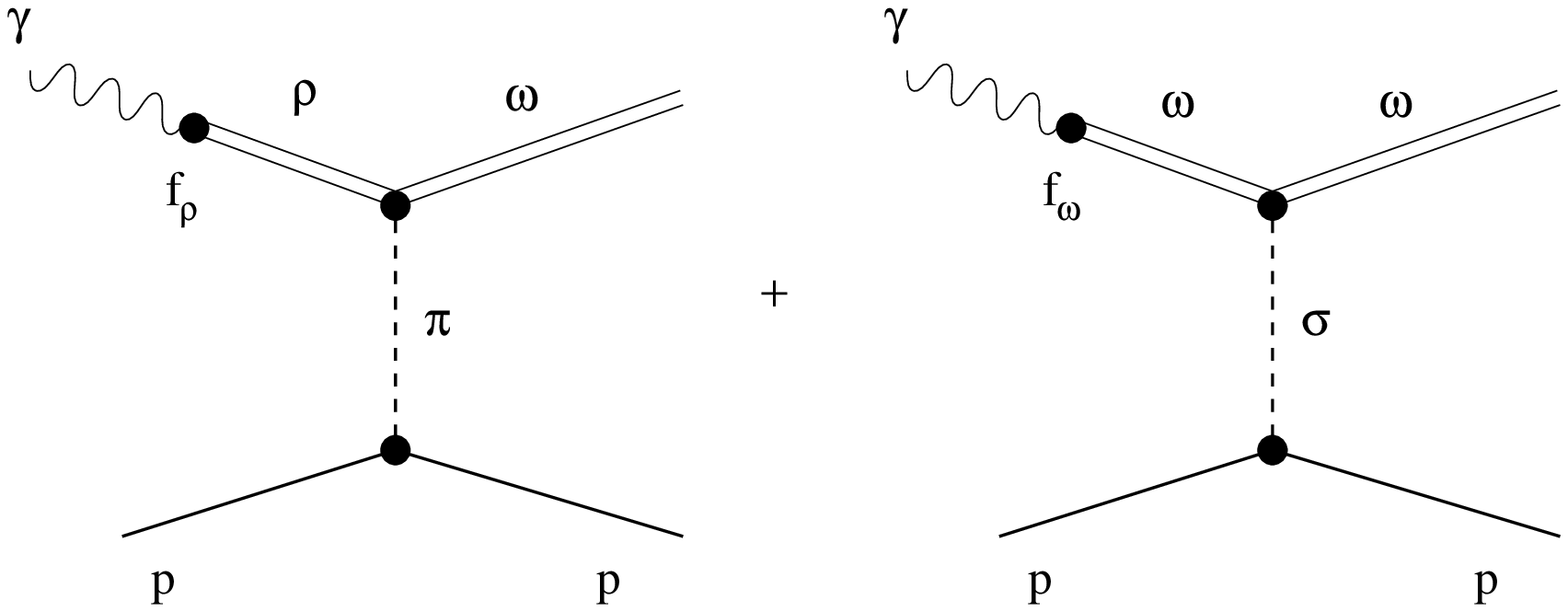,
height=85mm,angle=-90}}
\end{picture}

Fig.~2 t-channel exchange contributions to the $\gamma p \rightarrow
\omega p$ reaction using the current-field identities of Ref. [18].

\newpage
\setlength{\unitlength}{1mm}
\begin{picture}(150,60)
\put(10,0){\epsfig{file=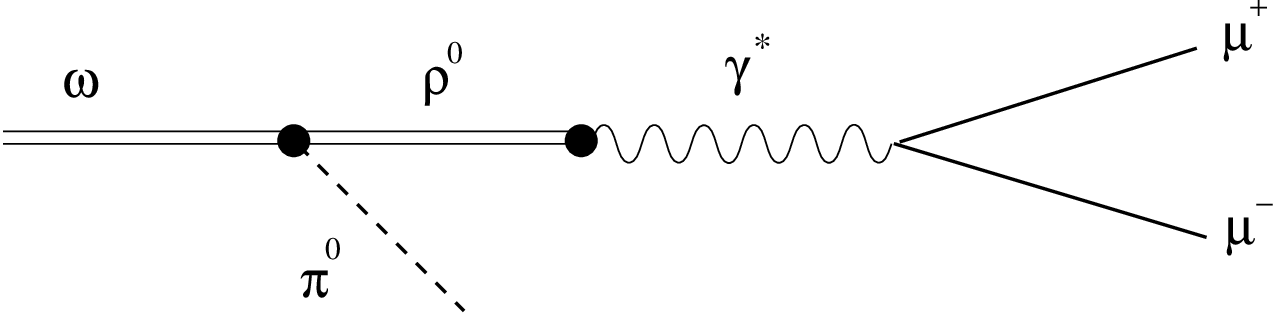,
height=30mm,angle=-90}}
\end{picture}

Fig.~3 Vector Dominance Model of the $\omega$ Dalitz decay form factor.

\setlength{\unitlength}{1mm}
\begin{picture}(150,140)
\put(0,10){\epsfig{file=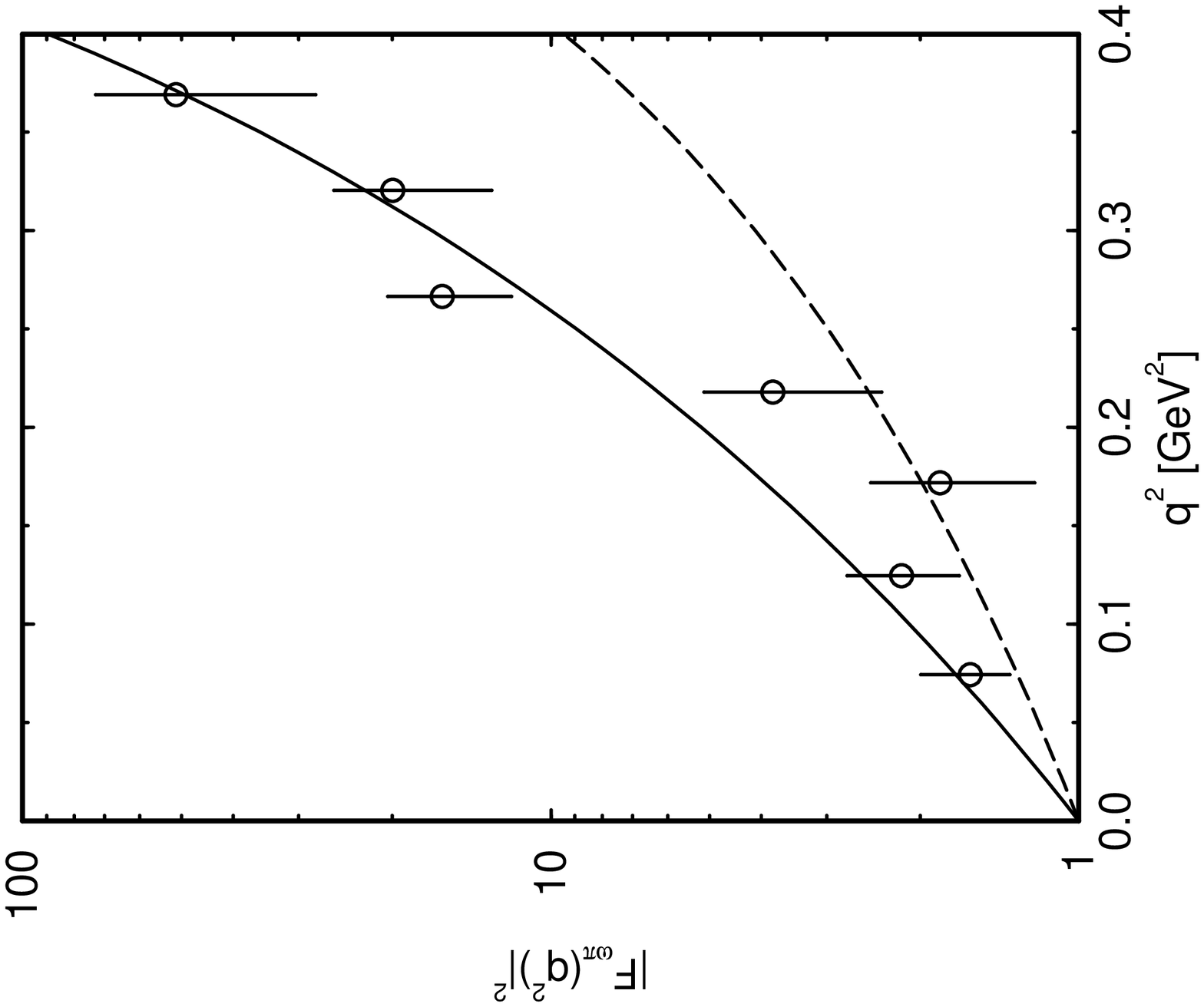,
height=130mm,angle=-90}}
\end{picture}

Fig.~4 The data on the $\omega$ Dalitz decay form factor [19,20]
compared to the Vector Dominance Model of Fig. 3 (dashed line)
and to the same model with a nonlocal $\pi\rho\, \omega$
vertex of the form ${M^2_\rho}/{(M^2_\rho - q^2)}$ (full line).

\newpage
\setlength{\unitlength}{1mm}
\begin{picture}(150,90)
\put(0,0){\epsfig{file=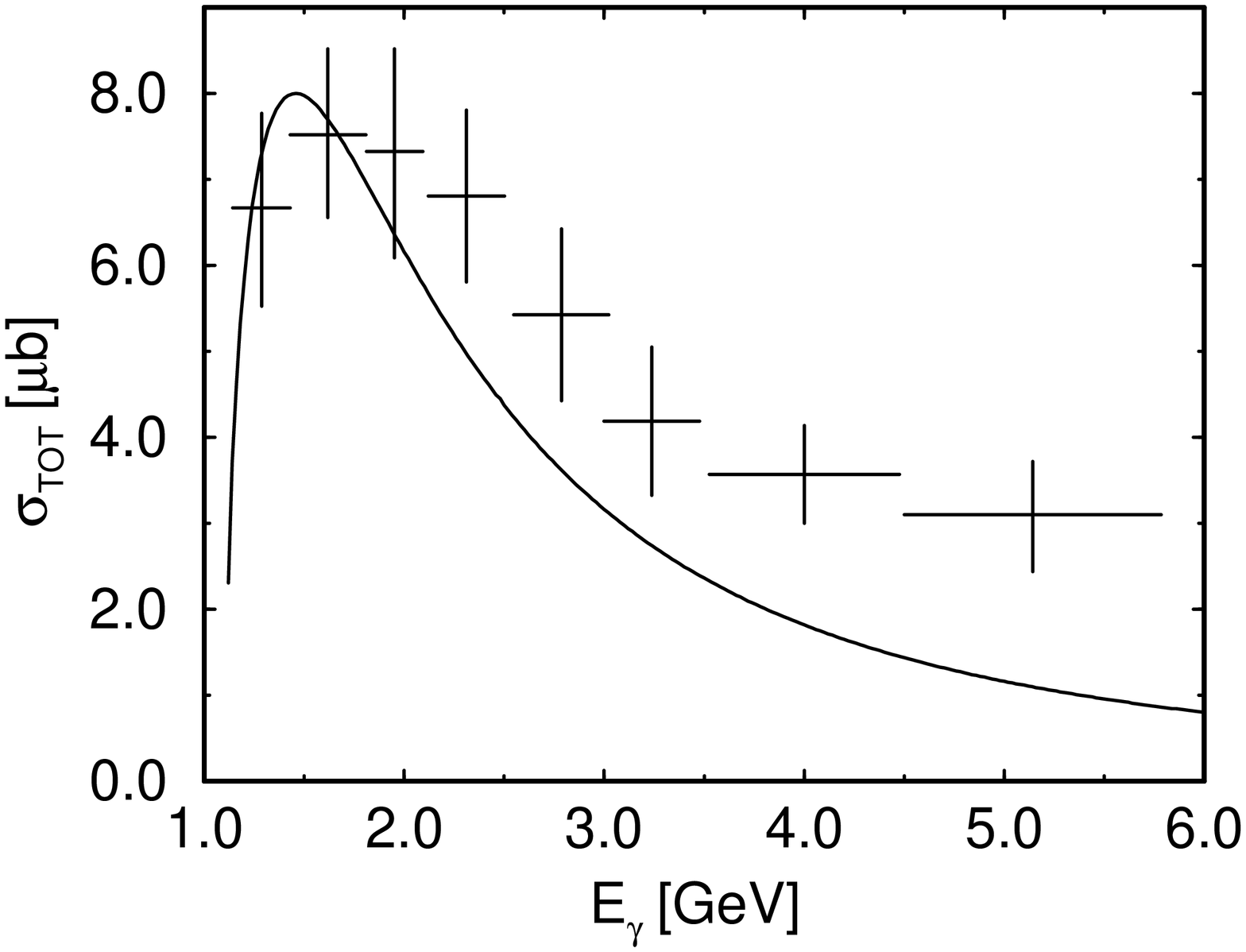,
height=100mm,angle=-90}}
\end{picture}

Fig.~5 Energy dependence of the $\pi$-exchange model of the $\gamma p
\rightarrow \omega p$ total cross section. The data are from
Ref. [11], where the meaning of the different sets of data points
is explained.

\setlength{\unitlength}{1mm}
\begin{picture}(150,80)
\put(0,0){\epsfig{file=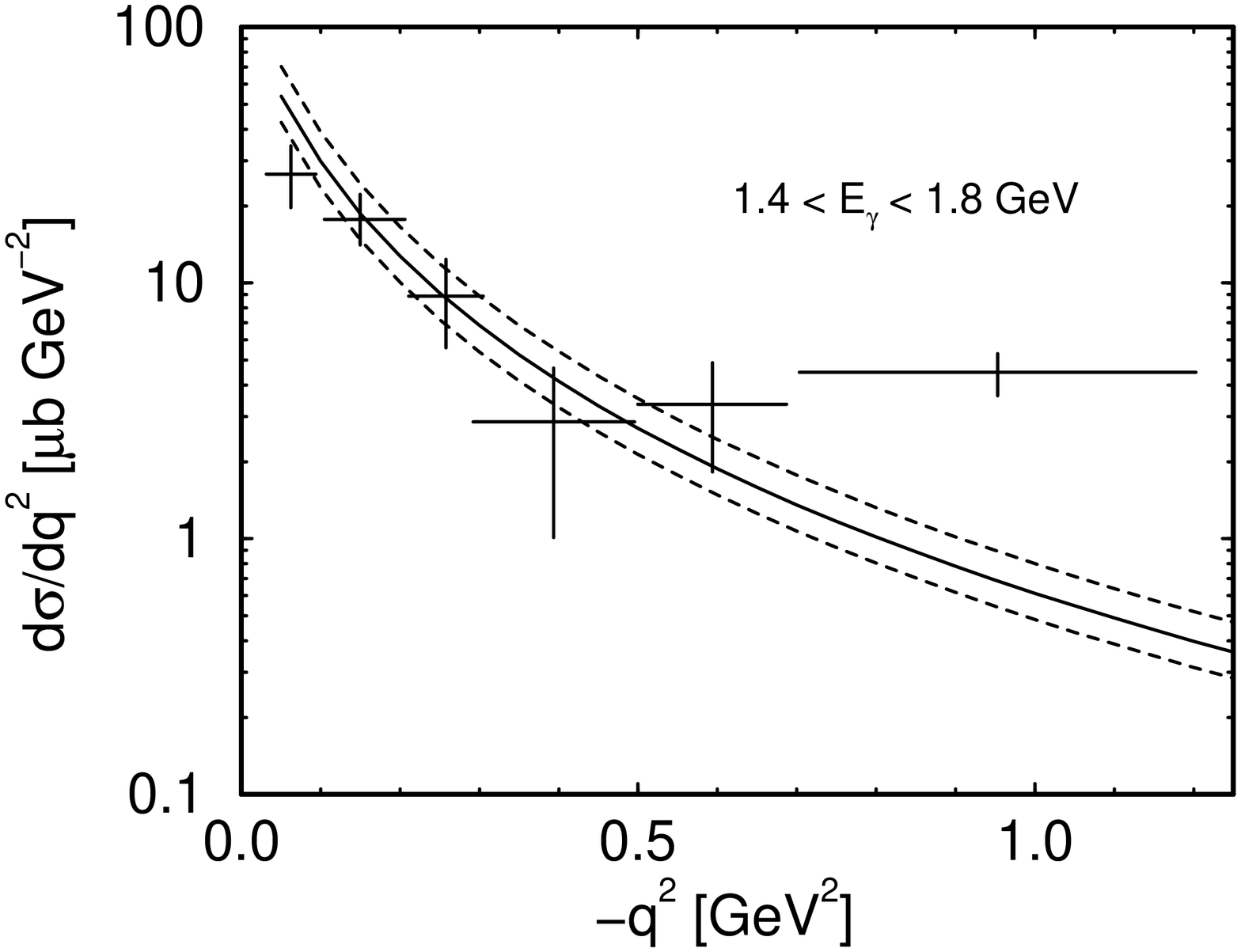,
height=100mm,angle=-90}}
\end{picture}

Fig.~6 Differential cross section $d \sigma/dq^2$ for the $\gamma p
\rightarrow \omega p$ reaction at 1.4 $< E_\gamma<$ 1.6 GeV.
The data are from Ref. [11]. The full
line is the $\pi$-exchange model for $E\gamma$ = 1.6 GeV. The dotted
lines show the uncertainties due to the photon energy resolution

\begin{picture}(150,80)
\put(-10,-20){\epsfig{file=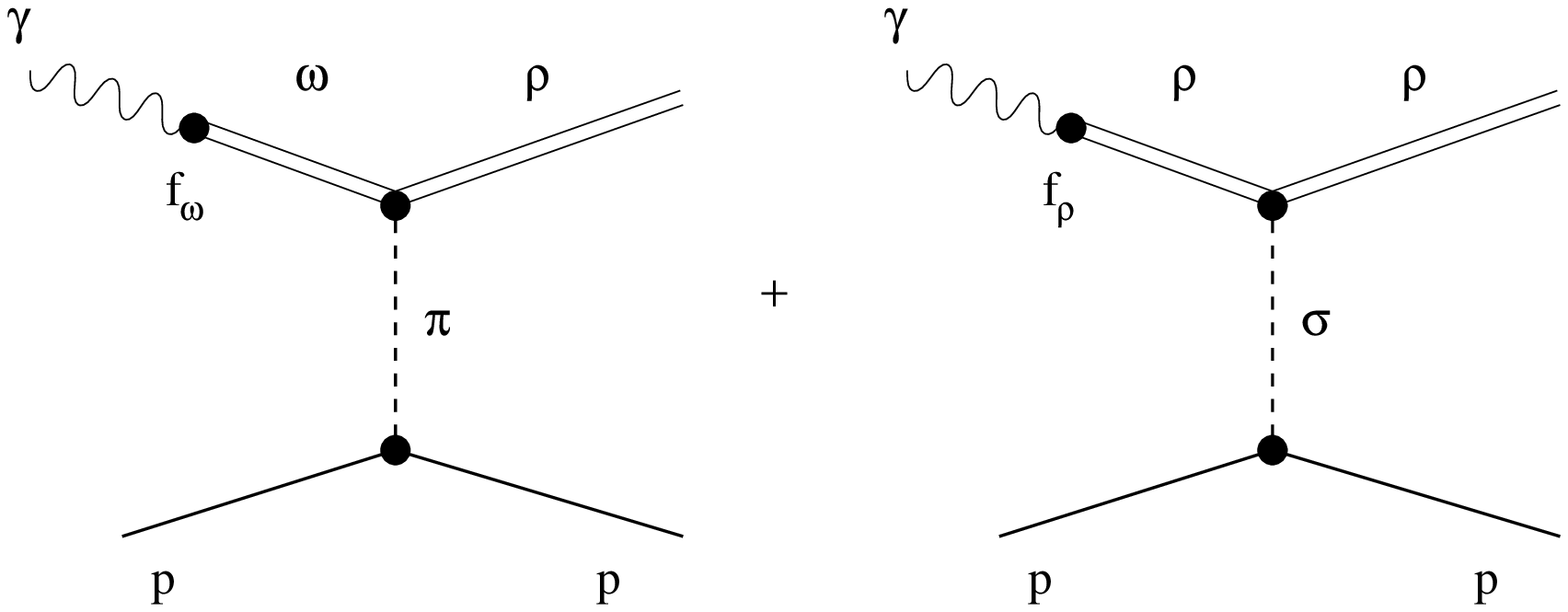,
height=85mm,angle=-90}}
\end{picture}

Fig.~7 t-channel exchange contributions to the $\gamma p \rightarrow
\rho^0 p$ reaction.

\begin{picture}(150,80)
\put(0,0){\epsfig{file=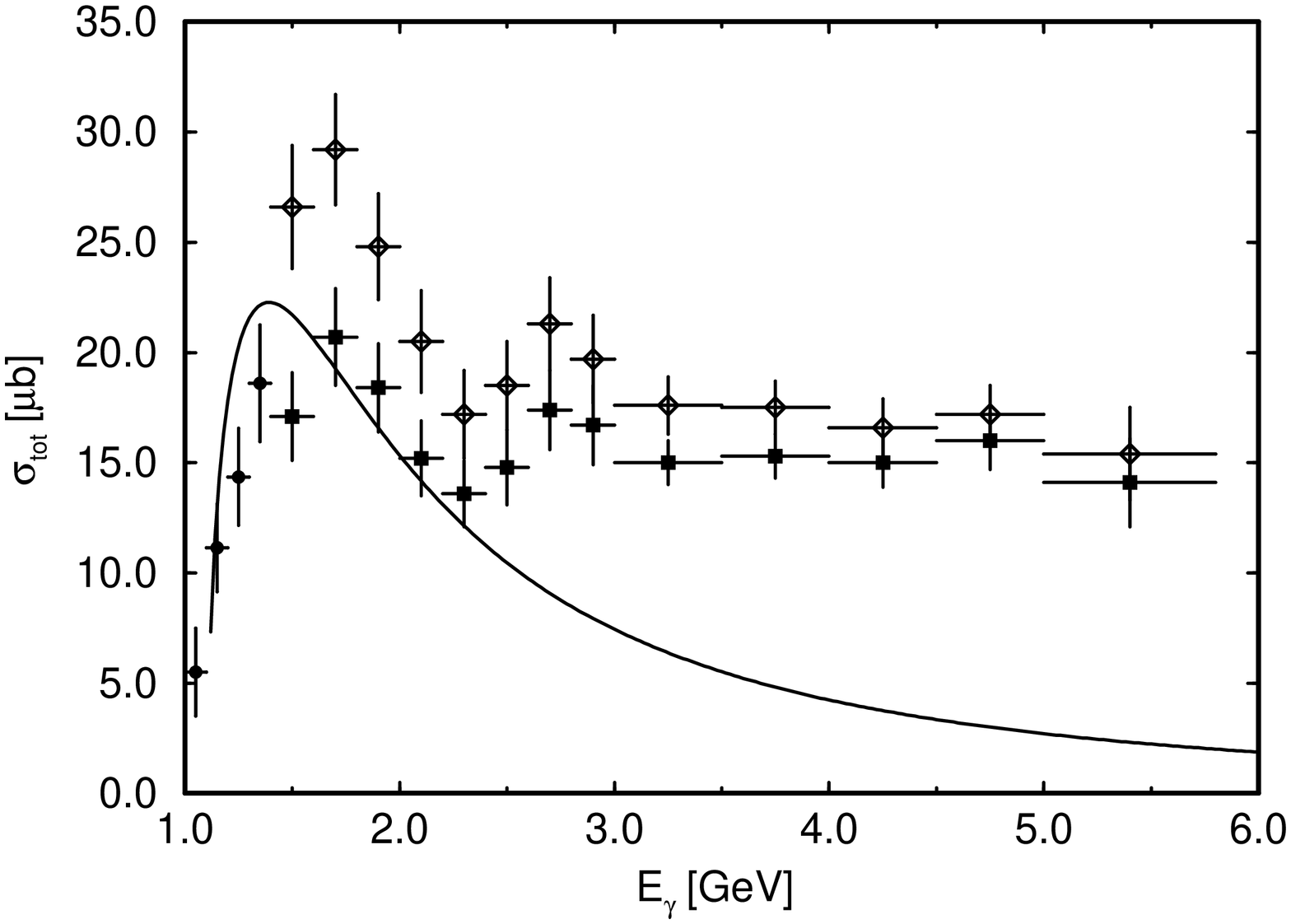,
height=100mm,angle=-90}}
\end{picture}

Fig.~8 Energy dependence of the $(\pi + \sigma)$-exchange model of
the $\gamma p \rightarrow \rho^0 p$ total cross section. The data are
from Ref.[11]

\setlength{\unitlength}{1mm}
\begin{picture}(150,80)
\put(0,-10){\epsfig{file=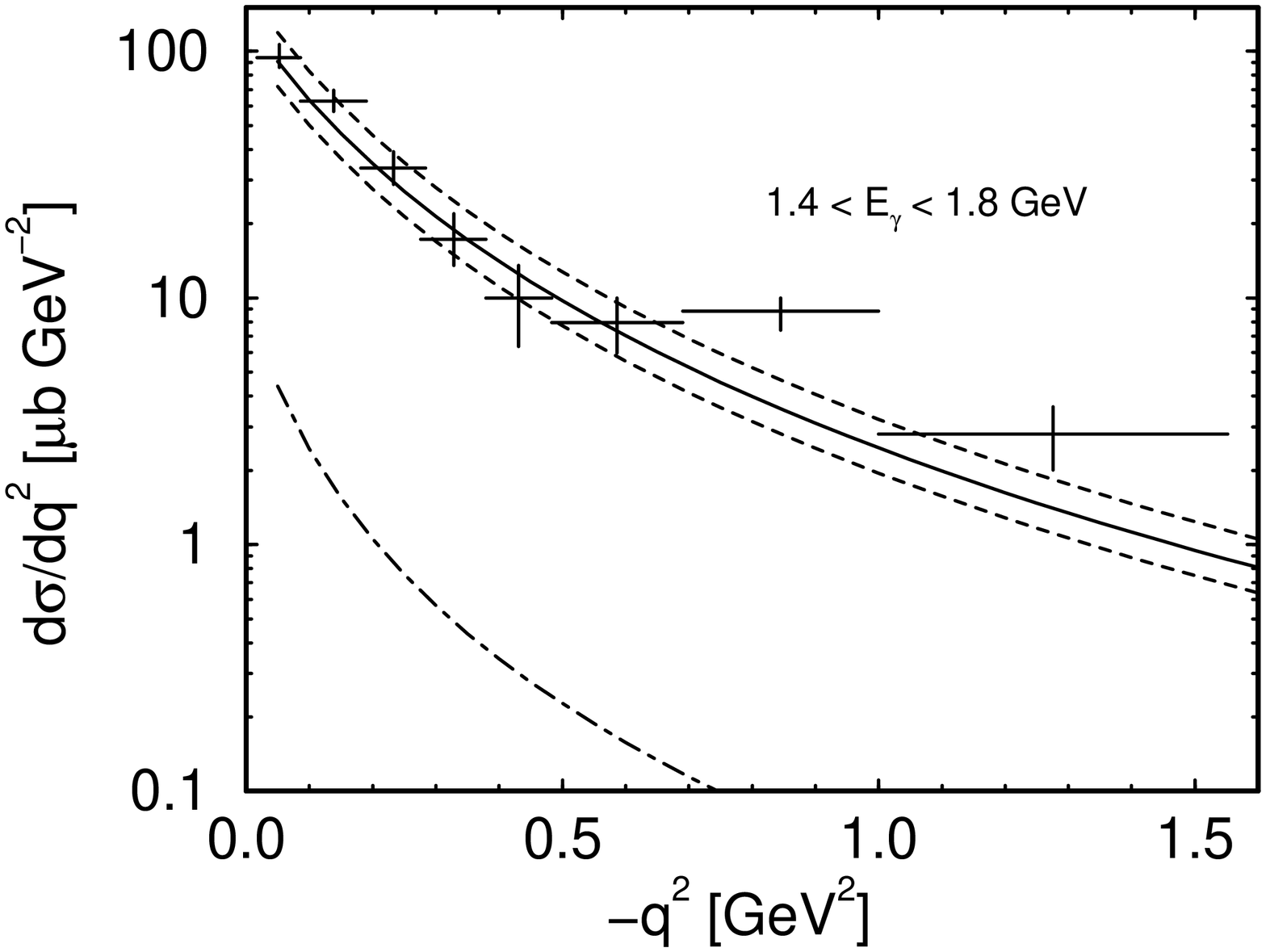,
height=100mm,angle=-90}}
\end{picture}

Fig.~9 Differential cross section $d\sigma/dq^2$
for the $\gamma p \rightarrow \rho^0 p$
reaction. The data
are from Ref. [11]. The full line is the $(\pi + \sigma)$-exchange
model for $E_\gamma$ = 1.6 GeV. The dot-dashed line shows the
$\pi$-exchange contribution. The dotted lines show the uncertainties due
to the photon energy resolution.

\setlength{\unitlength}{1mm}
\begin{picture}(150,65)
\put(20,0){\epsfig{file=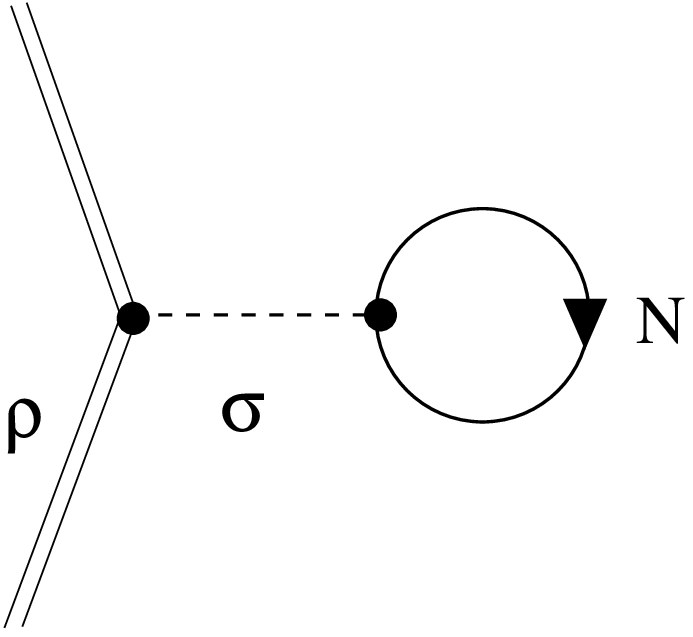,
height=60mm,angle=-90}}
\end{picture}

Fig.~10 The $\rho$-meson self-energy in matter to lowest order in the density.

\setlength{\unitlength}{1mm}
\begin{picture}(150,120)
\put(5,0){\epsfig{file=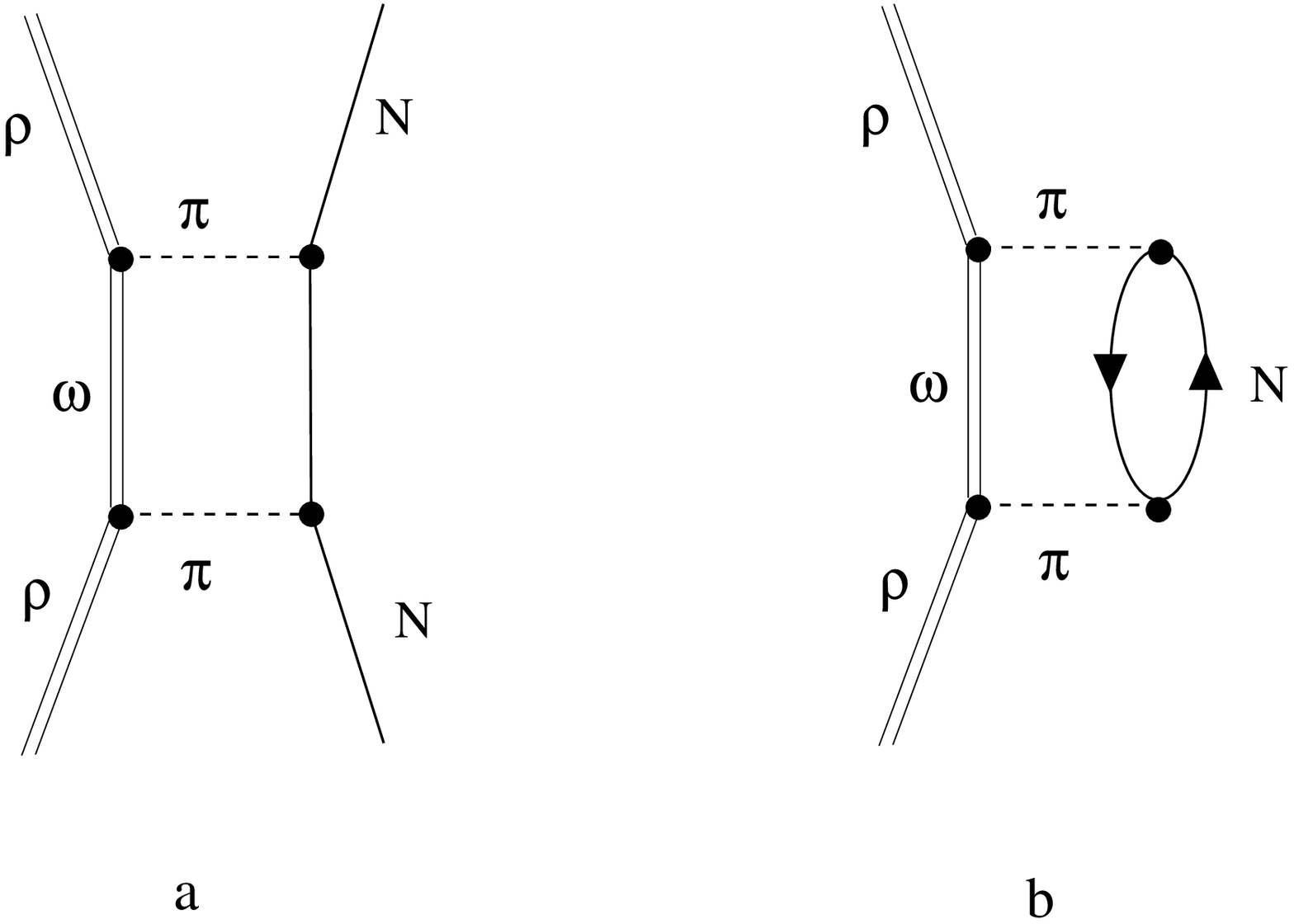,
height=100mm,angle=-90}}
\end{picture}

Fig.~11 The $2\pi$-exchange $\rho$-nucleon interaction (a) and the
corresponding $\rho$-meson self-energy in matter (b).

\end{document}